\documentclass[
preprint,
superscriptaddress,
 amsmath,amssymb,
 aps,
pra,
]{revtex4-2}

\usepackage[T1]{fontenc}
\usepackage[utf8]{inputenc}
\usepackage{setspace}
\usepackage{lineno}
\usepackage{mathrsfs}
\usepackage{geometry}
\usepackage[separate-uncertainty=true]{siunitx}
\usepackage{graphicx}% Include figure files
\usepackage{dcolumn}% Align table columns on decimal point
\usepackage{bm}% bold math
\usepackage{hyperref}
\hypersetup{colorlinks,allcolors=black}
\begin{document}

%\preprint{APS/123-QED}

\title{Electro-optically tunable single-frequency lasing  from neodymium-doped lithium niobate microresonators}% Force line breaks with \\

\author{Yannick Minet}
\email{yannick.minet@imtek.uni-freiburg.de}
    \affiliation{
Laboratory for Optical Systems, Department of Microsystems Engineering - IMTEK, University of Freiburg, Georges-K{\"o}hler-Allee 102, 79110 Freiburg, Germany}
 \affiliation{Gisela and Erwin Sick Chair of Micro-Optics, Department of Microsystems Engineering - IMTEK, University of Freiburg, Georges-K{\"o}hler-Allee 102, 79110 Freiburg, Germany}

\author{Simon J. Herr}
\affiliation{Fraunhofer Institute for Physical Measurement Techniques IPM, Georges-K{\"o}hler-Allee 301, 79110 Freiburg, Germany\\}
\author{Ingo Breunig}
\email{ingo.breunig@ipm.fraunhofer.de}
    \affiliation{
Laboratory for Optical Systems, Department of Microsystems Engineering - IMTEK, University of Freiburg, Georges-K{\"o}hler-Allee 102, 79110 Freiburg, Germany}
 \affiliation{Fraunhofer Institute for Physical Measurement Techniques IPM, Georges-K{\"o}hler-Allee 301, 79110 Freiburg, Germany\\}
\author{Hans Zappe}
    \affiliation{Gisela and Erwin Sick Chair of Micro-Optics, Department of Microsystems Engineering - IMTEK, University of Freiburg, Georges-K{\"o}hler-Allee 102, 79110 Freiburg, Germany}
\author{Karsten Buse}
    \affiliation{
Laboratory for Optical Systems, Department of Microsystems Engineering - IMTEK, University of Freiburg, Georges-K{\"o}hler-Allee 102, 79110 Freiburg, Germany}
 \affiliation{Fraunhofer Institute for Physical Measurement Techniques IPM, Georges-K{\"o}hler-Allee 301, 79110 Freiburg, Germany\\}

\date{\today}% It is always \today, today,

\begin{abstract}
Tunable light sources are a key enabling technology for many applications such as ranging, spectroscopy, optical coherence tomography, digital imaging and interferometry. For miniaturized laser devices, whispering gallery resonator lasers are a well-suited platform, offering low thresholds and small linewidths, however, many realizations suffer from the lack of reliable tuning. Rare-earth ion-doped lithium niobate offers a way to solve this issue.
Here we present a single-frequency laser based on a neodymium-doped lithium niobate whispering gallery mode resonator that is tuned via the linear electro-optic effect. Using a special geometry, we suppress higher-order transverse modes and hence ensure single-mode operation. With an applied voltage of just \SI{68}{\volt}, we achieve a tuning range of \SI{3.5}{\giga\hertz}. The lasing frequency can also be modulated with a triangular control signal. 
The freely running system provides a frequency and power stability of better than $\Delta\nu =\SI{20}{\MHz}$ and \SI{6}{\percent}, respectively, for a 30 minute period.  This concept is suitable for full integration with existing photonic platforms based on lithium niobate.
\end{abstract}

\maketitle

%\tableofcontents
%%%%%%%%%%%%%%%%%%%%%%%%%%  body  %%%%%%%%%%%%%%%%%%%%%%%%%%
\section{Introduction}
There is a strong demand for miniaturized, tunable lasers providing stable single-frequency lasing plus fast and wide tuning. Optical microresonators offer a possibility to accomplish this task since it is possible to fabricate them on on-chip material platforms. These microresonators are characterized by a high quality factor ($Q$-factor) and a small mode volume.
They can be fabricated with a large free spectral range, in the best case with only one resonator mode in the whole gain bandwidth of the laser medium. Additionally, they provide the possibility of designing the transvers mode spectrum which ensures single-mode lasing.

These properties can be used to passively stabilize lasers via self-injection locking \cite{Vassiliev98,Sprenger09,snigirev2021ultrafast,li2022integrated}.
In Ref. \cite{snigirev2021ultrafast}, a compact on-chip system with a small linewidth was realized. It can be linearly tuned as fast as \SI{12}{\peta\hertz\per\second}.
However, the tuning range of the system is limited by the locking bandwidth of the external resonator to the laser chip, which is about \SI{1}{\giga\hertz}.
In contrast to the self-injection technique, it is also possible to generate laser light directly from microresonators fabricated out of active laser media. The properties of microresonator described in the introduction before enable a low pump threshold and narrow linewidth. 
Such systems have been proposed and demonstrated in a large variety of material platforms \cite{KuwataGonokami98,He13}.
For example, ultra-low threshold lasing of a titanium doped sapphire whispering gallery laser has been shown recently \cite{Azeem.25.08.2021}.

Whereas single frequency operation and wafer-scale fabrication in most systems is possible, changing the laser frequency often remains challenging.
While approaches like temperature tuning or mechanical stress can achieve response times in the millisecond or microsecond range, linear electro-optic tuning offers even faster performance \cite{Werner17b,Padmaraju.2014}.
Lithium niobate (LN), as a non-centro-symmtetric crystal, offers the possibility of influencing the refractive index $n$ by the linear electro-optic effect, the Pockels effect. In addition, large non-linear coefficients and a wide transparency range make LN an attractive material for many applications \cite{Jia.2021,Guenter2012,Lin.2020,Zhu:21}.

Additionally, rare-earth doped lithium niobate can serve as a laser material platform.
A wide range of active components has been experimentally demonstrated on this platform, such as amplifiers and lasers \cite{Huang.1996,Bruske.2017,Bruske.2019,Wang.2021,Yin.2021, Sohler91b, Sohler2013}.
The outstanding optical performance as well as the possibility of doping LN makes it a promising candidate to realize monolithic tunable laser sources.
In Ref. \cite{Yin.2021} an electro-optically tunable erbium-doped lithium niobate microring laser has been demonstrated.
However, the laser possesses multi-mode operation during the tuning and the electro-optical tuning efficiency suffers from the small spatial overlap between the optical mode and the electric control field.
Furthermore, the electro-optic tuning behavior of such a microresonator laser has never been analyzed in detail.

In this work, we demonstrate an electro-optically tunable single-frequency laser based on a neodymium-doped lithium niobate microresonator.
Using a four-level system as provided by the trivalent neodymium ions has the advantage of leading to a far lower lasing threshold compared with that of a three-level system. 
In order to ensure a high excitation efficiency we resonantly pump the whispering-gallery resonator (WGR).
Using extraordinarily polarized light, i.e. polarized parallel to the $c$-axis, has two advantages:
Firstly, the absorption cross-section of the Nd-ions for the pump light is higher than that for ordinarily polarized light \cite{Fan86}, resulting in a more efficient excitation scheme. Secondly, for electro-optic tuning of the lasing frequency we can make use of the Pockels coefficient $r_{33}$ which is in our case the largest accessible electro-optic (EO) coefficient of lithium niobate, requiring lower voltages.

To ensure single-mode operation, we make use of a so-called photonic belt resonator geometry,  which suppresses higher-order tranverse modes \cite{Grudinin.2015b}.
By this micrometer-sized structure, we can also prevent undesired self-frequency doubling given by fortuitous phase-matching with higher order modes \cite{Herr17b}.
In the following, we first elucidate the theoretical quantities such as the pump threshold and the influence of the Pockels effect on the resonance frequency. Then, we experimentally verify the single-frequency operation and analyze the stability of the laser. Finally the tuning of this laser light source is demonstrated.

\section{Theoretical considerations}
The threshold power $P_\mathrm{th}$ for lasing in a microresonator can be determined by \cite{Herr17},
\begin{linenomath}
\begin{align}
    P_\mathrm{th}= \frac{2\pi n_\mathrm{l}h c_0 V_\mathrm{p}}{\lambda_\mathrm{l} \lambda_\mathrm{p} Q_\mathrm{l} \sigma_\mathrm{em} \tau \eta}
    \label{Pthres}
\end{align}
\end{linenomath}
with the refractive index at the lasing wavelength $ n_\mathrm{l}$, the mode volume of the pump mode $V_\mathrm{p}$, the lasing wavelength $\lambda_\mathrm{l}$, the pump wavelength $\lambda_\mathrm{p}$, the quality factor at the lasing wavelength $Q_\mathrm{l}$, the emission cross section $\sigma_\mathrm{em}$, the lifetime $\tau$ and the excitation efficiency $\eta$. 
In Ref. \cite{Herr17} $\eta$ was derived for a non-resonant pump scheme. We will in the following deduce $\eta$ for our resonant pump scheme.
The parameter $\eta$ is defined as the fraction of the absorbed pump power in order to pump the laser process $P_\mathrm{abs}$ over the total pump power $P$ as
\begin{linenomath}
\begin{align}
		\eta &= \frac{P_\mathrm{abs}}{P}
		=\frac{P_\mathrm{int}}{P}(1-\mathrm{e}^{-\alpha L})
		\label{Pabs_vs_Pin}
\end{align}
\end{linenomath}
with the absorption coefficient $\alpha$, the path length $L=2\pi R$, the radius $R$ and the internal power $P_\mathrm{int}$. We assume that absorption is the dominant loss at the pump wavelength and every absorbed photon leads to the excitation of the upper laser level.
Furthermore, we use the a first order approximation for $\mathrm{e}^{-\alpha L}\approx1-\alpha L$.
Expressing the resonant power enhancement $P_\mathrm{abs}/P$ by the  intrinsic finesse of the resonator $\mathscr{F}_\mathrm{int}$ and the coupling efficiency $K$, Eq. (\ref{Pabs_vs_Pin}) reads as
\begin{linenomath}
\begin{align}
\eta=\frac{\mathscr{F}_\mathrm{int}}{2\pi}K\alpha L
\label{power_enhancement}
\end{align}
\end{linenomath}
Assuming that the pump light is non-detuned from the resonance of the pump mode and inserting the intrinsic finesse as $\mathscr{F}_\mathrm{int}=2\pi/\alpha L$ into Eq. (\ref{power_enhancement}), we obtain :
\begin{linenomath}
\begin{align}
\eta=K
\label{etagleichK}
\end{align}
\end{linenomath}
Consequently,the highest excitation efficiency and therefore the lowest threshold is observed in critical coupling with $K=1$. %However, due to the strong absorption of the pump light it is likely that critical coupling can not be achieved.

Let us consider e-polarized pump light with $\lambda_\mathrm{p}=\SI{813}{\nano\meter}$.
With $\sigma_\mathrm{em}=\SI{1.8e-19}{\centi\meter\squared}$ \cite{Fan86}, $\tau=\SI{100}{\micro\second}$, $ V_\mathrm{p}=\SI{e6}{\micro\meter\cubed}$, $n=2.15$, $Q_\mathrm{l}=6\times10^6$ and $\eta=1$ we obtain $P_\mathrm{th}\approx\SI{0.3}{\milli\watt}$.

We can also estimate the introduced shift of the lasing frequency $\nu_\mathrm{l}$ by an external electric field $E_z$.
The resonance frequency of the WGR, being identical with the possible laser frequency $ \nu_\mathrm{l}$, can be estimated by \cite{Breunig16}
\begin{linenomath}
\begin{equation}
    \nu_\mathrm{l}\approx m\frac{c}{2\pi R n_\mathrm{l}}
    \label{eq:eiegnfrequency}
\end{equation}
\end{linenomath}
for the major radius of the microresonator $R$, the azimuthal mode number $m$, the vacuum speed of light $c$ and the  refractive index of the bulk material $n_\mathrm{l}$ at the lasing wavelength.

From Eq. (\ref{eq:eiegnfrequency}) it is clear that either changing the radius $R$ or the refractive index $n_l$ will change the eigenfrequency.
Since lithium niobate posses the Pockels effect, the lasing frequency can be changed by applying an external voltage to the crystal. The refractive index is modified according to
\begin{linenomath}
\begin{align}
	\Delta n_\mathrm{l} = -\frac{1}{2}  n_\mathrm{l}^3 r E_z.
	\label{eq:pockels_n}
\end{align}
\end{linenomath}
for $r$ the Pockels coefficient and $ E_\mathrm{z} =U_\mathrm{c}/d$ the  electric field resulting from the control voltage $U_\mathrm{c}$ divided by the thickness $d$ of the WGR.
Combining Eqs.\,(\ref{eq:eiegnfrequency})  and (\ref{eq:pockels_n}),  the change of the eigenfrequency $\Delta\nu_\mathrm{l}$  can then be expressed as
\begin{linenomath}
\begin{align}
	\Delta\nu_\mathrm{l}  = \frac{1}{2}  n_\mathrm{l}^2 r E_z \nu_\mathrm{l}.
	\label{eq:pockels_frequency}
\end{align}
\end{linenomath}
We neglect the contribution of the piezo-electric effect to the frequency change since it is an order of magnitude smaller than the Pockels effect \cite{weis1985lithium}.
% ?
Using Eq.\,(\ref{eq:pockels_frequency}) we can make an estimate for the expected frequency tuning range. Assuming $U_\mathrm{c}=\SI{62.5}{\volt}$ and a thickness $d$ of \SI{250}{\micro\meter} which results in a electric field strength of $E=\SI{250}{\volt\per\milli\meter}$, a Pockels coefficient $r_{33}=\SI{25}{\pico\meter\per\volt}$ \cite{mendez1999wavelength} and a lasing frequency of $\nu_\mathrm{l}=\SI{276.15}{\tera\hertz}$, we determine the frequency shift to be $\Delta\nu\approx\SI{4.0}{\giga\hertz}$.
%DONE!
% Does everyone automatically know what this wedge geometry looks like? If not, the following is pretty unclear.
Since we employ a special, wedge-like geometry (Fig. \ref{fig:setup} \textbf{a})) we need to consider a reduction of the electric field compared of that of a simple plate capacitor.
It has been shown before, that the overlap between the externally applied electric field and the optical electric field is reduced depending on the size of the belt \cite{Minet.2021b}. A drop of about \SI{50}{\percent} for larger sizes of the wedge is expected. Additionally, with larger sizes the homogeneity of the electric control field decreases.
However, a small size of wedge is desired to suppress higher-order modes and therefore ensure single mode lasing and suppress undesired self-frequency doubling \cite{Herr17b}.

\section{Experimental methods}
The microresonator was fabricated from a commercially available single-domain congruent $z$-cut lithium niobate (LiNbO$_3$) wafer of \SI{250}{\micro\meter} doped with nominally \SI{0.4}{\mole\percent} neodymium (Nd)
and \SI{5.1}{\mole\percent} magnesium oxide (MgO) (Yamaju Ceramics Co. Ltd.).
To provide a low resistance electrical connection, the $+z$ and $-z$-sides of the wafer are coated with a 250-nm-chrome layer.
To cut a resonator blank out of the wafer, we use a femtosecond laser with a repetition rate of \SI{60}{\kilo\hertz} and a wavelength of \SI{343}{\nano\meter}. The average power was \SI{1}{\watt}.
To ensure a sufficient mechanical and electrical connection, the crystal blank is soldered to a brass post using low temperature soldering tin.

\begin{figure}[ht]
\centering\includegraphics{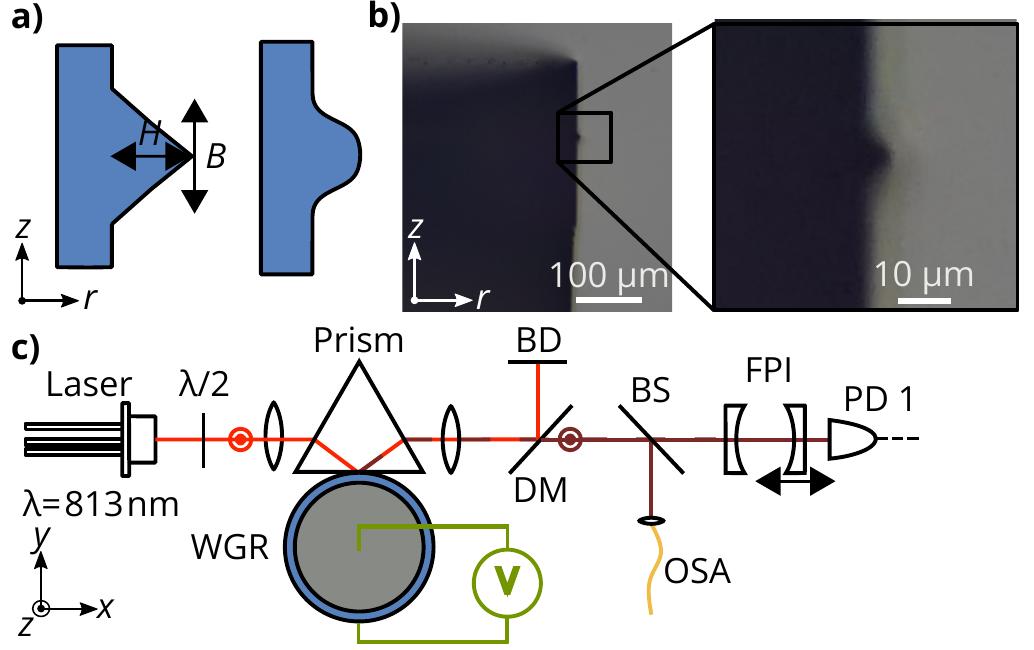}
\caption{
\textbf{a)} Drawing of the resonator with the corresponding coordinates of the base $B$ and the height $H$ after laser cutting (left) and polishing (right).
\textbf{b)} Microscopic side view of the fabricated wedge-like structure of the resonator after polishing.
\textbf{c)} A schematic of the measurement setup. The laser with $\lambda_\mathrm{l}=\SI{813}{\nano\meter}$ is coupled via a rutile prism to the whispering gallery resonator (WGR). The polarization is controlled by a $\lambda/2$-wave-plate. The remaining pump light is filtered using a dichroic mirror (DM) and sent to a beam dump (BD).
The lasing beam is split by a 50:50 beam splitter (BS) and further analyzed using a scanning Fabry-P\'erot-Interferometer (FPI) and an optical spectrum analyzer (OSA).
\label{fig:setup}}
\end{figure}

Then, using the same femtosecond laser, we shape the resonator using a computer-controlled lathe.
The WGR has a major radius $R=\SI{800}{\micro\meter}$, resulting in a free spectral range (FSR) of \SI{27.1}{\giga\hertz}. The optic axis is parallel to the axis of symmetry of the resonator.
The wedge-like geometry, shown in Fig.\,\ref{fig:setup} a), has a base $B$ of \SI{7.5}{\micro\meter}  and a height $H$ of \SI{5}{\micro\meter}.
In the final step of the machining in order to improve the quality factor of the resonator, a single run of \SI{2}{\min} chemical mechanical polishing is done. As shown in Fig. 1 b), after polishing, the previously manufactured corners of the wedge are slightly rounded off.

The setup for the optical characterization consists of a pump laser, coupling optics and mechanics, a voltage source for the tuning, a protective housing and devices to characterize the properties of the emitted beam (Fig.\,\ref{fig:setup} c)). As a pump laser we use either a titanium sapphire laser or a $813$-nm single frequency laser diode. While both lasers are used for characterization, only the Ti:Sa laser is used as a pump laser in the electro-optical tuning experiments. The extraordinarily polarized pump beam is coupled resonantly into the resonator via a rutile prism.

The residual pump light and the generated laser light are separated using a dichroic mirror and sent to a beam dump.
The generated laser light is split into two arms to be further analyzed either using a scanning Fabry-P\'erot-Interferometer (FPI) or an optical spectrum analyzer (OSA).
The FPI has a FSR of \SI{1.5}{\giga\hertz} and is scanned by multiple FSRs during the experiment. The linewidth of the FPI is \SI{10}{\mega\hertz}. 
The OSA (Yokogawa AQ6370D ) is used to measure the absolute frequency and to resolve a larger frequency range. It has a resolution of \SI{15}{\giga\hertz}.
To maintain similar conditions for all experiments, the WGR is housed in an acrylic glass cover. Measurements begin after a period of a few minutes to allow the system to attain thermal stability, since no temperature controller is active for the experiments.
For the electro-optic control of the resonance frequency a copper wire is bonded onto the top of the WGR while the bottom is connected to the brass post.
The control signal from the function generator is amplified by a 10 volt-per-volt high\textminus voltage amplifier with a maximum output of \SI{150}{\volt}.
Using this setup, we measure the linewidth the WGR at $\lambda=\SI{813}{\nano\meter}$ as well as close to the expected lasing wavelength; the derived quality factors at pump and lasing wavelengths are $Q_\mathrm{p}=3\times 10^5 $ and $Q_\mathrm{l}=6\times 10^6 $, respectively.
From a spectroscopic measurement  we determine an absorbance of \SI{0.6}{\per\cm} at $\lambda=\SI{813}{\nano\meter}$. The resulting intrinsic quality factor based on this value is $Q= 2.8\times10^5$.

\section{Results and discussion}
When the resonator is excited with extra-ordinarily polarized light at only $P_\mathrm{p}= \SI{4\pm0.5}{\milli\watt}$ laser excitation is clearly visible and emission of \SI{1086}{\nano\meter} wavelength light is observed in both emission directions.
This power value is higher than the estimated before, the difference stemming mainly from the unknown prism-to-resonator coupling strength during the laser operation.

%Moreover, a geometrical mismatch in the overlap of the pump light and the pump mode would also increase the lasing threshold.

\begin{figure}[ht]
\centering\includegraphics{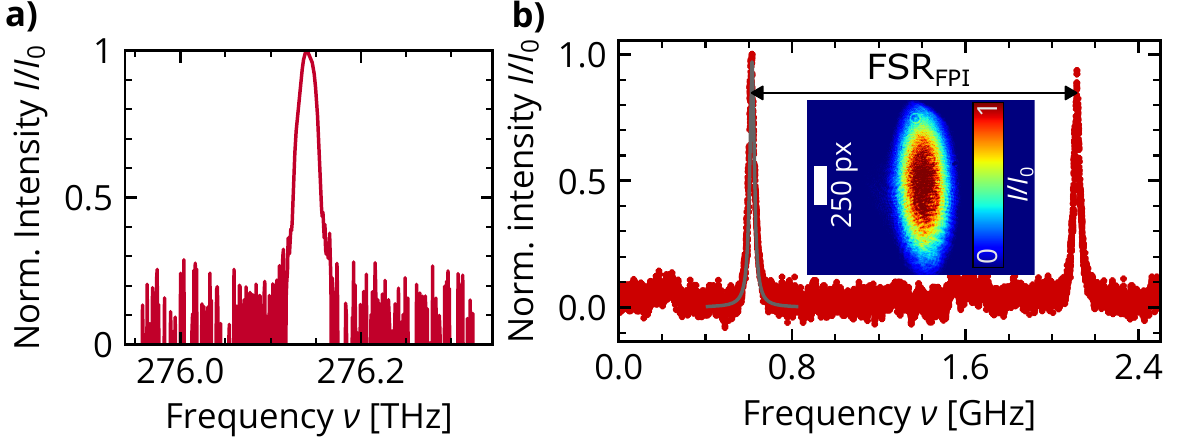}
\caption{Characterization of the emitted laser light. \textbf{a)} The recorded spectra of the optical spectrum analyzer (OSA) and in \textbf{b)} of the scanning Fabry-P\'erot-Interferometer (FPI). The inset in \textbf{b)} shows a false-color picture recorded with a beam profile camera.}
\label{fig:FPI}
\end{figure}

As can be seen in Fig. \ref{fig:FPI} a) we observe only a single emission peak at the expected frequency $\nu_\mathrm{l}=\SI{276.15}{\tera\hertz}$; the entire measurement covers a frequency range of \SI{450}{\giga\hertz}.
The peak position corresponds to $\lambda_\mathrm{l}= \SI{1085.6}{\nano\meter}$.

For higher spectral resolution we launch the generated laser light into the scanning FPI. The recorded spectrum can be seen in Fig. \ref{fig:FPI} b).
Since only one peak is visible in the free spectral range, added to the fact that we scan over the entire 430-GHz gain bandwidth of Nd:MgO:LN \cite{MacKinnon:94},
we verify single frequency operation.
By fitting a Lorentzian function to the transmission peak, we determine a linewidth of the transmitted laser light similar to the linewidth of the FPI of \SI{10}{\mega\hertz}, indicating that the measurement is equipment-limited.

The inset in Fig. \ref{fig:FPI} b) shows a false-color image of the beam profile, taken after collimating the beam with a 40-mm bi-convex lens. It shows a TEM$_{00}$-like shape, another clear indication of single mode operation.

\begin{figure}[ht]
\centering\includegraphics{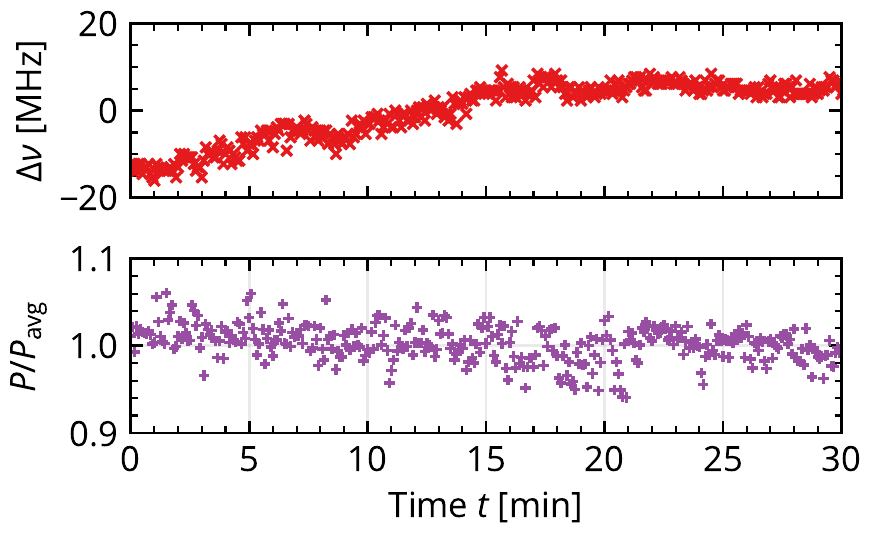}
\caption{Stability of the emission frequency and power of the free running system. The frequency was recorded by tracking the peak position of the scanning Fabry-P\'erot-Interferometer (FPI). The power of the lasing is measured using an additional photodiode placed behind a 50:50 beam splitter in front of the FPI.}
\label{fig:stab}
\end{figure}

For measurement of power and frequency stability, an additional beam splitter and a photodiode are placed in front of the FPI.
First, we couple about \SI{12}{\milli\watt} of pump light into the resonator, leading to a lasing power of $P_\mathrm{l}=\SI{340}{\micro\watt}$.  
The result is shown in Fig. \ref{fig:stab} where the upper part shows the frequency stability and the lower one, the power stability for a \SI{30}{\min} period.
The lasing frequency $\nu_\mathrm{l}$ shows a drift of about $\SI{20}{\mega\hertz}$ and the output power of the laser fluctuates within $\pm$ \SI{6}{\percent} with a slight downward slope.
Since the setup is not actively temperature stabilized, we assume that the frequency drift is caused by a change of the ambient temperature.

To demonstrate wavelength tuning, we apply a voltage that increases linearly in \SI{10}{\second} to a maximum value of \SI{70}{\volt}.
Such a slow ramp ensures that the resonator can thermalize at the in-coupled power and therefore remains in thermal equilibrium.
The shift in peak position on the scanning FPI while increasing the voltage results in a measure for the change of the lasing frequency $\nu_\mathrm{l}$ as shown in Fig. 4).
The maximum frequency shift we could obtain was \SI{3.5}{\giga\hertz}. If we increase the voltage further we cannot observe lasing anymore. It is likely that the pump mode frequency is then shifted so far from resonance that we decouple the resonator from the pump light.

\begin{figure}[ht]
\centering\includegraphics{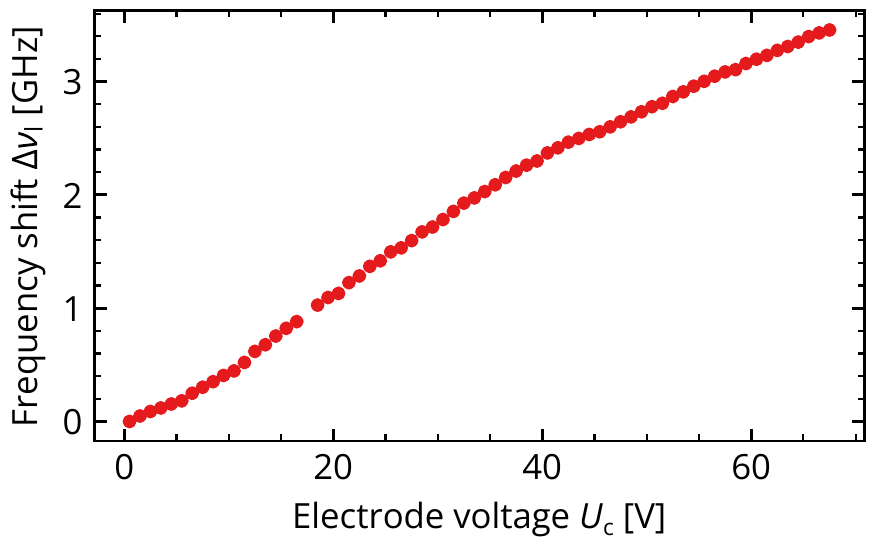}
\caption{Electro-optic tuning of the WGR-laser: Frequency shift $\Delta \nu_\mathrm{l}$ of lasing frequency vs. applied voltage.}
\label{fig:slowtuning}
\end{figure}

The magnitude of the observed lasing frequency shift corresponds to the GHz-broad resonances of the WGR, due to the strong absorption of the pump light.
We observe a non-linear tuning behavior with increasing voltage. As we use a pump power of several \si{\milli\watt}, it is likely that thermal effects influence the frequency tuning characteristic. A change of the resonator temperature of just \SI{10}{\milli\kelvin} leads to change of the resonance frequency $\Delta\nu \approx \SI{100}{\mega\hertz}$ \cite{Gorodetsky06}.
Further investigations are needed to understand the origin of this nonlinearity. 

The frequency shift amounts \SI{3.5}{\giga\hertz} at \SI{67.5}{\volt}.
This is lower than the theoretically expected value of \SI{4.3}{\giga\hertz} if the full electric field would be present.
However, a reduction is expected because of the wedge-like geometry of the resonator \cite{Minet.2021b} and if we assume a reasonable reduction of \SI{15}{\percent}, the theoretical value matches the experimentally measured one.

\begin{figure}[ht]
\centering\includegraphics{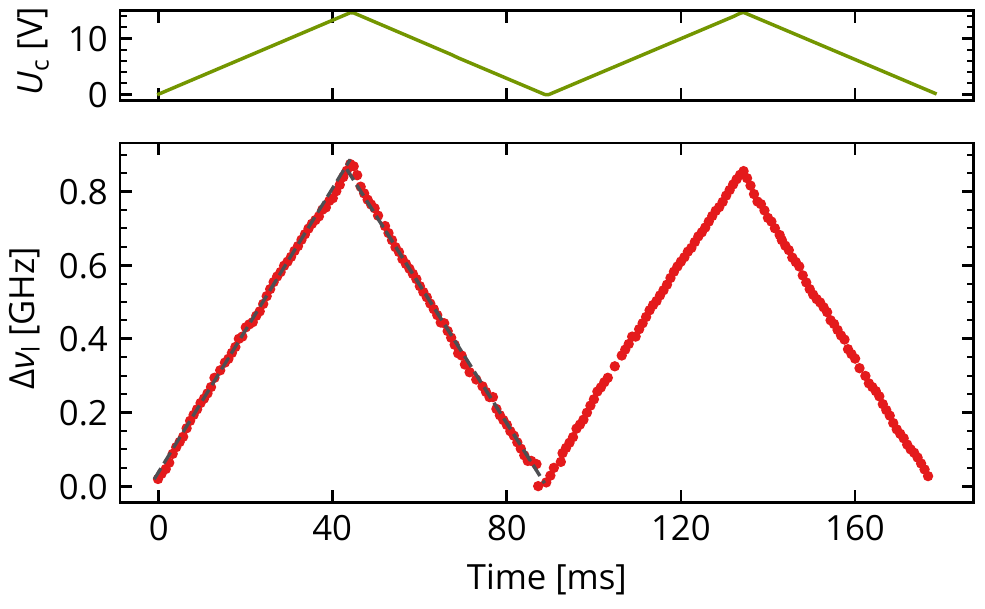}
\caption{Electro-optic tuning of the WGR-laser on the millisecond time scale. In the top plot we show the applied control voltage $U_\mathrm{c}$ on the electrodes. In the bottom the recorded frequency shift $\Delta\nu_\mathrm{l}$ is displayed. The dash-dotted lines are linear fits to the recorded data.}
\label{fig:fasttuning}
\end{figure}

Finally, we quantify the tunability of the presented system for smaller tuning ranges but for higher tuning speeds. To do so, we apply a triangular voltage sweep to the resonator with a frequency of \SI{11.25}{\hertz}.
The electric control signal and the resulting frequency shift can be seen in Fig. \ref{fig:fasttuning}).
By applying \SI{15}{\volt} to the electrode we observe a frequency shift $\Delta\nu=\SI{0.85}{\giga\hertz}$ on the rising side  resulting in a tuning rate of $\SI{57}{\mega\hertz\per\volt}$. One the rising side of the first triangle  we extract from the linear fit $R^2= 0.99$ and on the falling side $R^2= 0.95$, indicating a high linearity of the observed frequency shift. 

\section{Conclusion and Outlook}
In summary, we implement an electro-optically tunable laser based on a Nd:Mg:LN microresonator, showing stable single-frequency lasing centered around \SI{1086}{\nano\meter} wavelength with a TEM$_{00}$-like beam profile. The lasing frequency can subsequently be tuned more than \SI{3.5}{\giga\hertz} without mode-hops.
This free-running WGR laser provides \SI{6}{\percent} output power stability as well as $\SI{20}{\mega\hertz}$ frequency stability without any feedback system for a period of \SI{30}{\min}. 
We have also demonstrated that using a triangular control signal leads to a corresponding linear shift of the lasing frequency, making the system useful as a possible light source for ranging and sensing, for example.

Since lithium niobate can be doped with other rare-earth ions such as praseodymium, tunable lasing could also be realized at visible wavelengths

A limitation of our demonstration is that resonant pumping limits the stable tuning range to the width of the pump mode. Furthermore, resonant pumping affects the tuning behavior, because less light is coupled into the resonator when a voltage is applied,
which leads to a reduction in temperature and is most likely the origin of the nonlinear tuning behavior when the lasing frequency is tuned wide and slowly. One approach to tackle this issue is to employ a side-pumping scheme \cite{Herr.2019}. By doing so, even an inexpensive pump source like a 813-nm multi-mode laser diode can be employed. This ensures that the same pump power is always present during the electro-optic tuning. However, the quality factors achieved here were too low to use these schemes efficiently.
Realizing this approach on an on-chip platform would push another limit: a smaller radius would lead to a larger FSR and hence larger tuning ranges would not be impeded by mode-hops, when only one longitudinal mode is present in complete gain bandwidth of the laser material \cite{Herr.2019}. Furthermore, by implementing this scheme in integrated photonic platforms based on lithium niobate will lead to an increase of the electric field strength using the same voltage as used in this demonstration by two orders of magnitude and thus will allow for wider tuning - we consider this work to be an important step towards inexpensive tunable light sources.

\begin{acknowledgments}
Y.M. acknowledges financial support by a Gisela and Erwin Sick Fellowship.
The authors thank Jan Szabados and Nicolas Amiune for fruitful discussions on the manuscript.
\end{acknowledgments}

\bibliographystyle{unsrt}
\bibliography{EO_WGR_LASER}% Produces the bibliography via BibTeX.

\end{document}